\documentclass[sigconf]{acmart}
\AtBeginDocument{%
  }

\usepackage{pgfplots}
\pgfplotsset{compat=1.18}

\begin{document}

\title{Multi-Reward GRPO for Stable and Prosodic Single-Codebook TTS LLMs at Scale}

\author{Yicheng Zhong}
\email{ajaxzhong@tencent.com}
\orcid{0000-0001-7661-5384}
\affiliation{%
  \institution{Tencent Technology Co.Ltd}
  \city{Shenzhen}
  \state{Guangdong}
  \country{China}
}

\author{Peiji Yang}
\affiliation{%
  \institution{Tencent Technology Co.Ltd}
  \city{Shenzhen}
  \state{Guangdong}
  \country{China}
}
\email{peijiyang@tencent.com}

\author{Zhisheng Wang}
\affiliation{%
  \institution{Tencent Technology Co.Ltd}
  \city{Shenzhen}
  \state{Guangdong}
  \country{China}
}
\email{plorywang@tencent.com}

\renewcommand{\shortauthors}{Yicheng et al.}

\begin{abstract}
Recent advances in Large Language Models (LLMs) have transformed text-to-speech (TTS) synthesis, inspiring autoregressive frameworks that represent speech as sequences of discrete codec tokens. Among them, single-codebook TTS LLMs have emerged as compact and streamable architectures that jointly model semantic and acoustic integration. However, despite their efficiency, these models often exhibit unstable prosody, speaker drift, and degraded naturalness. To address these issues, we propose a multi-reward Group Relative Policy Optimization (GRPO) framework that directly optimizes the token generation policy of single-codebook TTS LLMs. Beyond standard intelligibility and speaker similarity objectives, our design integrates three rule-based rewards: a length penalty for duration consistency, an entropy regularization reward for decoding stability, and an LLM-annotated prosody alignment reward that explicitly supervises rhythm. In this prosody reward, an external reasoning LLM predicts multiple plausible pause structures via in-context learning, providing a human-preference-aligned supervisory signal for GRPO training. To assess universality, we further attach a flow-matching (FM) decoder on top of the GRPO-optimized AR backbone and observe consistent additional gains, indicating that our reinforcement optimization enhances the intrinsic AR policy. We further conduct a scalability analysis across data sizes and model scales, revealing that the proposed method consistently enhances prosodic stability, speaker similarity, and overall speech naturalness in single-codebook TTS LLMs.
\end{abstract}




\maketitle

\section{Introduction}
\begin{figure*}[!htbp]
  \includegraphics[width=\textwidth]{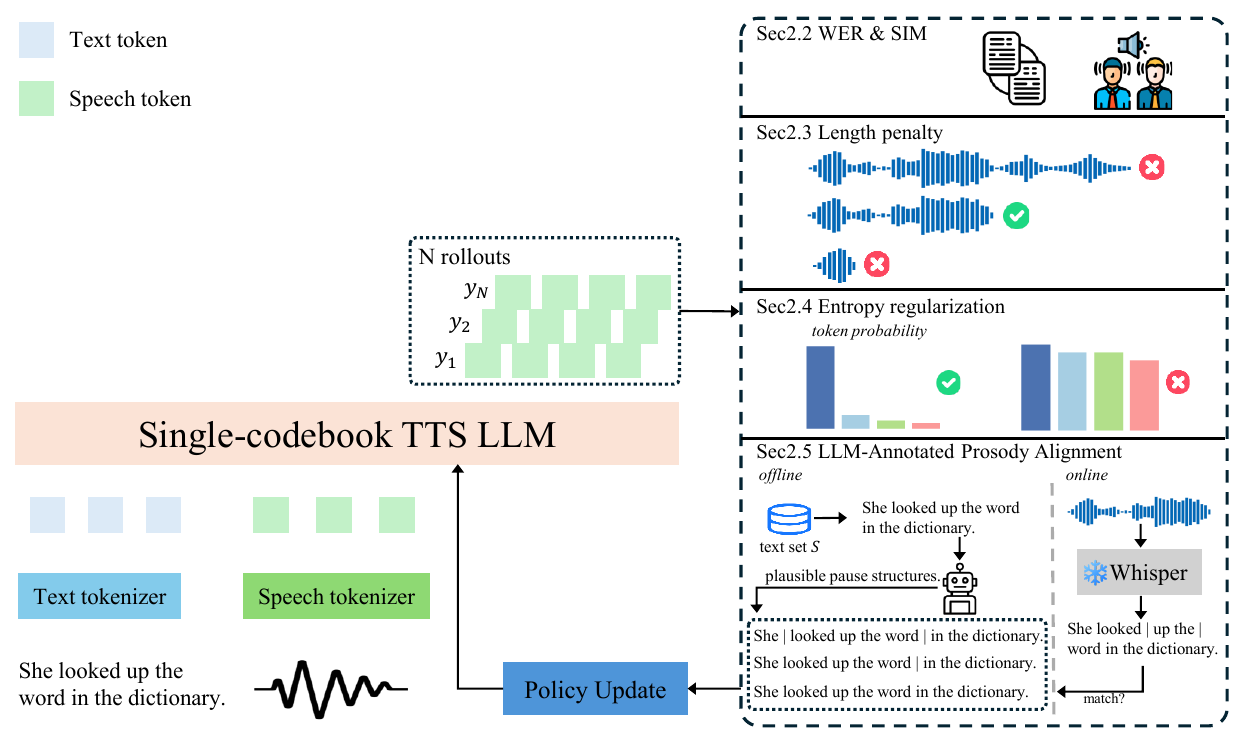}
  \caption{Overview of our framework.
Multiple rollouts are evaluated with WER/SIM, length penalty, entropy regularization, and an LLM-annotated prosody alignment reward to guide policy updates.}
  \label{fig:method}
\end{figure*}
Recent advances in Large Language Models (LLMs) have reshaped text-to-speech (TTS) synthesis, enabling autoregressive (AR) decoding over discrete codec tokens. Among these, single-codebook TTS LLMs stand out for their compactness and native streaming capability. Currently, zero-shot TTS systems span three main families: (1) LLM-based acoustic-token models with strong linguistic–acoustic modeling \cite{wang2025maskgct,wang2025spark}; (2) diffusion-based architectures that implicitly learn text–speech alignment with fine-grained control \cite{chen2025f5}; and (3) coarse-to-fine pipelines where AR LLMs predict semantic tokens refined by diffusion or flow-matching modules \cite{anastassiou2024seed,du2024cosyvoice,du2024cosyvoice2,du2025cosyvoice}. Within the single-codebook paradigm, semantic-focused systems rely on secondary models to recover acoustic detail, whereas joint semantic–acoustic systems directly generate tokens encoding both linguistic and paralinguistic cues \cite{ye2025llasa}. While the latter offers more expressivity and lower latency, unified modeling often introduces suboptimal decoding policies, leading to prosody instability, speaker drift, and weakened temporal controllability. Reinforcement learning (RL) provides a natural mechanism to directly optimize these AR policies but remains underexplored.

Prior RL efforts include uncertainty aware learning \cite{chen2024enhancing} and synthetic positive construction via reverse inference optimization \cite{hu2024robust}. Differentiable reward frameworks such as DiffRO \cite{gao2025differentiable} further enable supervised optimization with programmatic objectives, though DPO-style methods are sensitive to preference noise and costly to scale. In contrast, Group Relative Policy Optimization (GRPO) \cite{shao2024deepseekmath} stabilizes learning through group-wise advantage normalization and avoids dependence on dense preference labels.

We introduce a multi-reward GRPO framework that integrates objective metrics with three rule-based rewards: a length penalty for duration consistency, an entropy reward for stable decoding, and an LLM-annotated Prosody Alignment Reward for explicit rhythm supervision. To obtain prosodic templates, a reasoning LLM (e.g., DeepSeek-R1 \cite{guo2025deepseek}) generates multiple plausible pause patterns via in-context learning offline; these templates then guide online optimization toward natural rhythm. We further conduct a systematic scaling study across 1K–1M data and 1B–8B models, revealing a strong correlation between RL effectiveness and data scale. Experiments show robust gains in prosodic stability, speaker similarity, and naturalness. Moreover, adding a Flow Matching refinement module after RL continues to yield improvements, confirming that our method strengthens the intrinsic AR policy in a manner complementary to acoustic refiners.

\section{Methodology}
\subsection{Overview}
We propose a reinforcement learning framework based on Group Relative Policy Optimization (GRPO) to enhance the stability, prosody, and speaker similarity of single-codebook TTS large language models (LLMs). 
Starting from a single-codebook LLM as a policy $\pi_\theta(a_t|s_t)$, we employ GRPO to optimize the model with multiple complementary rewards that reflect both objective and rule-based criteria. 
The objective is to maximize the expected cumulative reward across the generated trajectory $\tau = (s_1, a_1, \ldots, s_T, a_T)$:
\begin{equation}
\mathcal{J}(\theta) = 
\mathbb{E}_{\tau \sim \pi_\theta}\left[
\sum_{t=1}^{T} R(s_t, a_t)
\right].
\end{equation}
The reward function $R$ is decomposed into several interpretable components that capture different aspects of synthesis quality:
\begin{equation}
R(s_t, a_t) = 
\alpha_{intl} R_{intl} +
\alpha_{sim} R_{sim} +
\alpha_{len} R_{len} +
\alpha_{ent} R_{ent} +
\alpha_{pro} R_{pro},
\label{eq:multi_reward}
\end{equation}
where $\alpha_i$ are tunable scaling coefficients that balance the contribution of different reward terms. The policy parameters $\theta$ are updated via GRPO to maximize $\mathcal{J}(\theta)$ under the combined signal.
\subsection{Intelligibility and Speaker Similarity Rewards}

\paragraph{Intelligibility Reward ($R_{intl}$).}
To measure intelligibility and fidelity, we use the pre-trained Whisper ASR model\cite{radford2022whisper} to transcribe the generated audio $A$ into text $\hat{S}$, and compute the Character Error Rate (CER) / Word Error Rate (WER) against the input text $S$:
\begin{equation}
R_{intl} = 1 - \frac{D_{lev}(\hat{S}, S)}{|S|},
\end{equation}
where $D_{lev}$ denotes the Levenshtein distance. 
This reward encourages the model to produce speech that is semantically consistent with the input text.

\paragraph{Speaker Similarity Reward}
To evaluate speaker consistency, we adopt the WavLM-large model\cite{chen2022wavlm} fine-tuned for speaker verification to extract speaker embeddings $E(A)$ and $E(A_{\text{ref}})$ from the generated and reference audios. 
We compute the cosine similarity between embeddings:
\begin{equation}
R_{sim} = \cos(E(A), E(A_{\text{ref}})) = 
\frac{E(A) \cdot E(A_{\text{ref}})}{\|E(A)\| \|E(A_{\text{ref}})\|}.
\end{equation}

\subsection{Length Penalty Reward}
To prevent premature stopping or overly long generations—a common instability issue in AR TTS—we constrain the generated speech duration $T$ relative to a target $T_{\text{target}}$. 
The target is estimated from the reference text length and the reference speech speed ratio. 
Given a tolerance range $[a, b]$, the reward is defined as:
\begin{equation}
R_{len} =
\begin{cases}
1, & \text{if } \frac{T_{\text{text}} / T}{r_{\text{ref}}} \in [a, b], \\
0, & \text{otherwise},
\end{cases}
\end{equation}
where $r_{\text{ref}}$ denotes the reference speaking rate computed from the paired reference input, which stabilizes the generation length distribution and mitigates truncation or excessive elongation.

\subsection{Entropy Regularization Reward}
To encourage stable and deterministic token generation, we regularize the policy entropy along the generation trajectory. 
Let $\bar{H}$ denote the average token-level entropy across the sequence, and $H_{\text{target}}$ be the entropy target estimated from high-quality samples. 
The entropy reward is formulated as:
\begin{equation}
R_{ent} = - \lambda_{ent} \cdot \max(0, \bar{H} - H_{\text{target}}).
\end{equation}
This reward penalizes excessively high entropy that leads to erratic prosodic variations, encouraging smoother generation paths.

\subsection{LLM-Annotated Prosody Alignment Reward}

\paragraph{Offline Annotation.}
To capture human-like prosody, we leverage an auxiliary reasoning LLM (\textit{DeepSeek-R1}) to annotate a set of input texts $S$ with appropriate pause structures. 
Few-shot examples of prosodic annotation are provided to the LLM for both Chinese and English. 
For Chinese, we adopt discrete pause markers (\#1–\#4) corresponding to increasing pause durations. 
For English, we employ \textit{Prosodic Word (PW)} and \textit{Prosodic Phrase (PPH)} labels, following the linguistic hierarchy in \cite{selkirk1980prosodic}. 
The resulting pseudo-labels serve as target prosodic structures.

\paragraph{Online Comparison.}
During GRPO training, the generated waveform is decoded and timestamped using Whisper. 
We convert silence durations into discrete pause symbols through a handcrafted rule-based mapping, and compare them against the pseudo-labels. 
If the predicted pause sequence matches the annotated pattern, a binary reward is assigned:
\begin{equation}
R_{pro} =
\begin{cases}
1, & \text{if } \hat{P}(A) \in \{P(S)\}, \\
0, & \text{otherwise},
\end{cases}
\end{equation}
where $\hat{P}(A)$ denotes the predicted pause structure from generated audio, and $\{P(S)\}$ is the annotated reference pattern set. 
This binary reward encourages alignment with human-preferred pause structures and suppresses unnatural rhythm patterns.

\section{Experiments}

\subsection{Implementation Details}
All experiments are conducted on \textbf{8$\times$H20 GPUs} with mixed-precision training. 
The GRPO optimization adopts a batch size of \textbf{16}, a learning rate of \textbf{1e--6}, and a group size of \textbf{12}. 
During generation, we employ the \texttt{vLLM} decoding configuration with 
\textit{top-k = 75}, \textit{top-p = 0.9}, \textit{temperature = 1.1}, and \textit{repetition penalty = 1.1}. 
The reward coefficients are empirically set as:
\[
\alpha_{\text{intl}} = \alpha_{\text{sim}} = \alpha_{\text{ent}} = \alpha_{\text{pro}} = 1.0, \quad 
\alpha_{\text{len}} = 0.1.
\]

For training data, we construct a bilingual corpus by sampling from Emilia\cite{he2024emilia} and libriheavy\cite{kang2024libriheavy}, 
selecting 1 million text samples (10--100 words) and 1 million speech samples balanced across Chinese and English (1:1 ratio). 
These are paired into approximately 1 million (\text{ref\_text}, \text{ref\_speech}, \text{target\_text}) triplets 
for online GRPO training, totaling about 5115 hours of audio.

\subsection{Main Results}
\begin{table}[!t]
  \centering
  \footnotesize
  \caption{Comparison of different methods on SEED test sets (test-zh, test-en). For each subset we report WER (↓ lower is better), SIM (↑ higher is better).}
  \label{tab:main-results}
  \begin{tabular}{lcccccccccc}
    \toprule
    Method          & \multicolumn{2}{c}{test-zh}         & \multicolumn{2}{c}{test-en}   & \multicolumn{2}{c}{test-hard}   &          \\
                    & CER$\downarrow$      & SIM$\uparrow$     & WER$\downarrow$      & SIM$\uparrow$         & CER$\downarrow$      & SIM$\uparrow$   & MOS$\uparrow$    \\
    \midrule
    Seed-TTS          & 1.12    &  \textbf{0.796}     &  2.62     &  \underline{0.714}  &  7.59    & \textbf{0.776}    & - \\
    FireRedTTS        & 1.51    &  0.635   &  3.82   &  0.460    &  17.45   & 0.621  & 3.53  \\
    MaskGCT           & 2.27    &  0.774   & 2.62   &  \underline{0.714}  &  10.27  &  0.748 & - \\
    F5-TTS            & 1.56   &   0.741   &  \textbf{1.83}  &  0.615   & 8.67   & 0.713 & 3.94 \\
    Spark-TTS         & 1.20   &  0.672  & \underline{1.98}  & 0.584  &  -   &  - & 4.01  \\
    CosyVoice        & 3.63   & 0.723   &  4.29   & 0.609   & 11.75  &  0.709 & 3.89 \\
    CosyVoice2        &  1.45  & 0.748  &  2.57  &  0.652  &  6.83  &  0.724 & 3.98 \\
    CosyVoice3      &  1.12  &  0.781  &  2.21  &  0.720  &  \textbf{5.83}  &  0.758 & 4.07 \\
    LLaSA-8B             & 1.59   &   0.684  &  2.97   & 0.574  &  11.09  &  0.660 & 3.67 \\
    \midrule
    LLaSA+SFT         &  1.51  &  0.688  & 2.89   &  0.582   &  10.63   & 0.674 & 3.76 \\
    LLaSA+RL        &  \underline{1.10}  &  0.758  &  2.12  &  0.672  &  6.04  & 0.731  & \underline{4.12} \\
    LLaSA+RL+FM    &  \textbf{1.08}   & \underline{0.790}  &  2.08  &  \textbf{0.733}  &  \underline{5.98}  & \underline{0.775} & \textbf{4.21} \\
    \bottomrule
  \end{tabular}
\end{table}



Table~\ref{tab:main-results} reports results on the SEED\cite{anastassiou2024seed} benchmark, evaluating both objective metrics (CER/WER, SIM) and subjective naturalness through MOS, where 100 randomly sampled utterances were rated by 10 participants. We compare against several systems including Seed-TTS \cite{anastassiou2024seed}, FireRedTTS \cite{guo2024fireredtts}, MaskGCT \cite{wang2025maskgct}, F5-TTS\cite{chen2025f5}, Spark-TTS\cite{wang2025spark}, CosyVoice\cite{du2024cosyvoice,du2024cosyvoice2,du2025cosyvoice}, and the LLaSA\cite{ye2025llasa} baseline, an SFT-only variant, and a post-RL Flow Matching (FM)\cite{mehta2024matcha} refinement model. Our GRPO-optimized LLaSA achieves the best CER on \textit{test-zh} and competitive results across languages, outperforming all open-source single-codebook TTS LLMs (e.g., Spark-TTS, CosyVoice/2) and the SFT-only counterpart, showing the higher sample efficiency of RL relative to supervised fine-tuning. Despite CosyVoice3 benefiting from a hybrid architecture and substantially larger training data (1M h vs.\ our 250k h), our model still attains lower CER and comparable SIM.

On the challenging \textit{test-hard} split, GRPO delivers notable gains in both CER and SIM, demonstrating improved robustness under difficult scenarios. Importantly, our method also achieves the highest MOS, indicating strong alignment with human preference. Adding FM after RL further improves performance, especially SIM and MOS, confirming that our RL optimization strengthens the intrinsic AR policy in a manner complementary to acoustic refinement.

\subsection{Scalability Analysis}
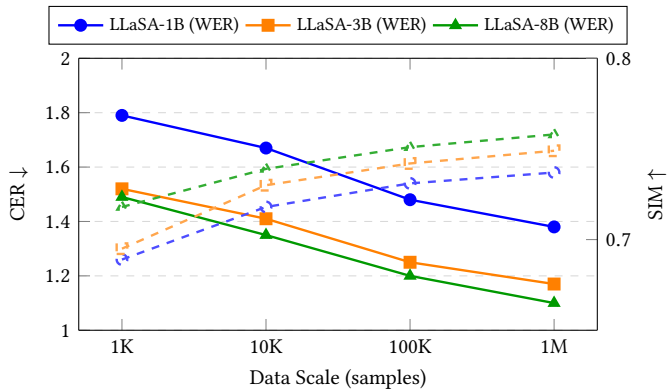
\begin{figure}[!t]
\centering
\begin{tikzpicture}
\begin{axis}[
    width=\columnwidth,
    height=5.2cm,
    xlabel={Data Scale (samples)},
    symbolic x coords={1K, 10K, 100K, 1M},
    xtick=data,
    axis y line*=left,
    ylabel={CER ↓},
    ymin=1, ymax=2,
    ymajorgrids=true,
    grid style={dashed, gray!30},
    tick label style={font=\small},
    label style={font=\small},
    legend style={at={(0.5,1.05)}, anchor=south, legend columns=-1, font=\footnotesize},
    mark size=2pt,
]

\addplot[line width=0.9pt, mark=*, color=blue]
    coordinates {(1K,1.79) (10K, 1.67) (100K,1.48) (1M,1.38)};
\addlegendentry{LLaSA-1B (WER)}

\addplot[line width=0.9pt, mark=square*, color=orange]
    coordinates {(1K,1.52) (10K, 1.41) (100K,1.25) (1M,1.17)};
\addlegendentry{LLaSA-3B (WER)}

\addplot[line width=0.9pt, mark=triangle*, color=green!60!black]
    coordinates {(1K,1.49) (10K, 1.35) (100K,1.20) (1M,1.10)};
\addlegendentry{LLaSA-8B (WER)}

\end{axis}

\begin{axis}[
    width=\columnwidth,
    height=5.2cm,
    symbolic x coords={1K, 10K, 100K, 1M},
    xtick=data,
    axis y line*=right,
    axis x line=none,
    ylabel={SIM ↑},
    ymin=0.65, ymax=0.8,
    ytick distance=0.1,
    ymajorgrids=false,
    tick label style={font=\small},
    label style={font=\small},
]

\addplot[dashed, line width=0.9pt, mark=o, color=blue!70]
    coordinates {(1K,0.689) (10K,0.718) (100K,0.731) (1M,0.737)};
\addplot[dashed, line width=0.9pt, mark=square, color=orange!70]
    coordinates {(1K,0.695) (10K,0.730) (100K,0.742) (1M,0.749)};
\addplot[dashed, line width=0.9pt, mark=triangle, color=green!60!black!80]
    coordinates {(1K,0.718) (10K,0.739) (100K,0.751) (1M,0.758)};

\end{axis}
\end{tikzpicture}
\caption{Scalability analysis.
WER (solid, left axis) decreases while SIM (dashed, right axis) increases as both data scale and model size grow.}
\label{fig:scaling}
\end{figure}
To investigate the effect of model and data scale, we apply the GRPO framework to 
models with 1B, 3B, and 8B parameters, trained with data scales 
ranging from 1K $\rightarrow$ 10K $\rightarrow$ 100K $\rightarrow$ 1M samples. 
Figure \ref{fig:scaling} visualizes the results, plotting \textit{CER} and \textit{SIM} against data scale.

We observe a clear monotonic trend: larger model capacities yield better prosodic stability and 
speaker preservation, while increasing GRPO data size further refines rhythm and decoding stability. 
Even small-scale RL (e.g., 10K samples) delivers measurable gains over the supervised baseline, 
highlighting the data efficiency of our proposed optimization scheme.
\subsection{Ablation Study}
\begin{table}[!t]
  \centering
  \small
  \caption{Ablation study on the contribution of different reward components. Lower CER/WER and higher SIM/MOS indicate better performance.}
  \label{tab:ablation}
  \begin{tabular}{lcccccc}
    \toprule
    Method & \multicolumn{3}{c}{zh}         & \multicolumn{3}{c}{en}   \\
    & CER$\downarrow$      & SIM$\uparrow$     & MOS$\uparrow$      & WER$\uparrow$         & SIM$\downarrow$      & MOS$\uparrow$        \\
    \midrule
    LLaSA & 1.59   &   0.684 & 3.68 &  2.97   & 0.574  &  3.57   \\
    + $R_{intl}$ \& $R_{sim}$  &  1.31  & 0.719 & 3.77 & 2.66 & 0.623  & 3.69  \\
    + $R_{len}$         & 1.23  &  0.738 & 3.81 & 2.48 & 0.647  &  3.75    \\
    + $R_{ent}$         & 1.12   & 0.751 & 4.01 & 2.25 & 0.668  &  3.90    \\
    + $R_{pro}$ (Full)  & 1.10  &  0.758 & 4.25 &  2.12  &  0.672  &  4.12  \\
    \bottomrule
  \end{tabular}
\end{table}
Table \ref{tab:ablation} reports the incremental effect of each reward term across both objective (CER/WER, SIM) and subjective (MOS) metrics. $R_{\text{intl}}$ \& $R_{\text{sim}}$ establish strong initial gains, while the length penalty further reduces duration mismatch. $R_{ent}$ yields notable improvements in both stability and MOS, indicating smoother and more natural token dynamics. $R_{pro}$ delivers the largest additional boost across all metrics—especially in MOS—showing that explicit rhythmic supervision not only improves prosodic structure but also better aligns the model’s outputs with human perceptual preferences. 
\section{Conclusion}
We introduced a GRPO-based RL framework that directly improves the intrinsic autoregressive policy of single-codebook TTS LLMs. By integrating objective, rule-based, and LLM-assisted prosody rewards, our method enhances prosodic stability, speaker similarity, and naturalness across model and data scales. These gains further compound with flow-matching refinement, indicating that our RL optimization strengthens the core AR decoder rather than overlapping with downstream token-refinement methods.
\bibliographystyle{ACM-Reference-Format}
\bibliography{ref}


\end{document}